\documentclass[aps,prl,twocolumn]{revtex4-2}
\usepackage{amssymb}
\usepackage{graphicx,amsmath}
\usepackage{hyperref}
\allowdisplaybreaks
\newcommand{\bea}{\begin{eqnarray}}
\newcommand{\eea}{\end{eqnarray}}
\newcommand{\beq}{\begin{equation}}
\newcommand{\eeq}{\end{equation}}
\newcommand{\benu}{\begin{enumerate}}
\newcommand{\enu}{\end{enumerate}}

\newcommand{\om}{\omega}

\newcommand{\ep}{\epsilon}

\newcommand{\prm}{\prime}

\newcommand{\bk}{{\bf k}}

\newcommand{\bv}{{\bf v}}

\usepackage{color}

\begin{document}

\title{
Exceptional van Hove Singularities in Pseudogapped Metals}
\date{\today}
\author{Indranil Paul$^1$ and Marcello Civelli$^2$}
\affiliation{
$^1$Universit\'{e} Paris Cit\'{e}, CNRS, Laboratoire Mat\'{e}riaux et Ph\'{e}nom\`{e}nes Quantiques,
75205 Paris, France\\
$^2$Universit\'{e} Paris-Saclay, CNRS, Laboratoire de Physique des Solides, 91405, Orsay, France
}

\begin{abstract}
Motivated by the pseudogap state of the cuprates, we introduce the concept of an ``exceptional''
van Hove singularity that appears when strong electron-electron interaction
splits an otherwise simply connected Fermi surface into multiply connected pieces. The singularity
describes the touching of two pieces of the split Fermi surface. We show that this singularity
is proximate to a second order van Hove singularity, which can be accessed
by tuning a dispersion parameter. We argue that, in a wide class of cuprates,
the end-point of the pseudogap is accessed only by
triggering the exceptional van Hove singularity. The resulting Lifshitz transition is characterized by
enhanced specific heat and nematic susceptibility, as seen in experiments.
\end{abstract}

\maketitle
\emph{Introduction.}---
In electronic systems on a lattice, the periodicity of the potential guarantees the existence of saddle points
of the dispersion $\ep_{\bk}$ as a function of the wavevector $\bk$ where the velocity
$\bv_{\bk} \equiv \nabla_{\bk} \ep_{\bk}$ vanishes~\cite{vanHove53,Ashcroft-Mermin}.
In two dimensions such van Hove singularities give
rise to diverging density of states, which has attracted attention since the early days of condensed matter
physics. Since electron-electron interaction is not needed to produce such singularities, they are typically
associated with non-interacting physics. The purpose of the current
work is to study ``exceptional'' van Hove singularities, which are saddle points
generated by strong electron-electron interaction. As we show below,
such a study is particularly useful to understand some unusual properties of
several hole-doped cuprates close to the doping where the pseudogap
disappears~\cite{cuprate-reviews,Klein19,Auvray19,Benhabib15,loret17,leyraud17,collignon21,michon21,
fang22,gourgout22,zhu22,kuspert22,girod21}.

\begin{figure}[b]
\begin{center}
\includegraphics[width=0.4\textwidth]{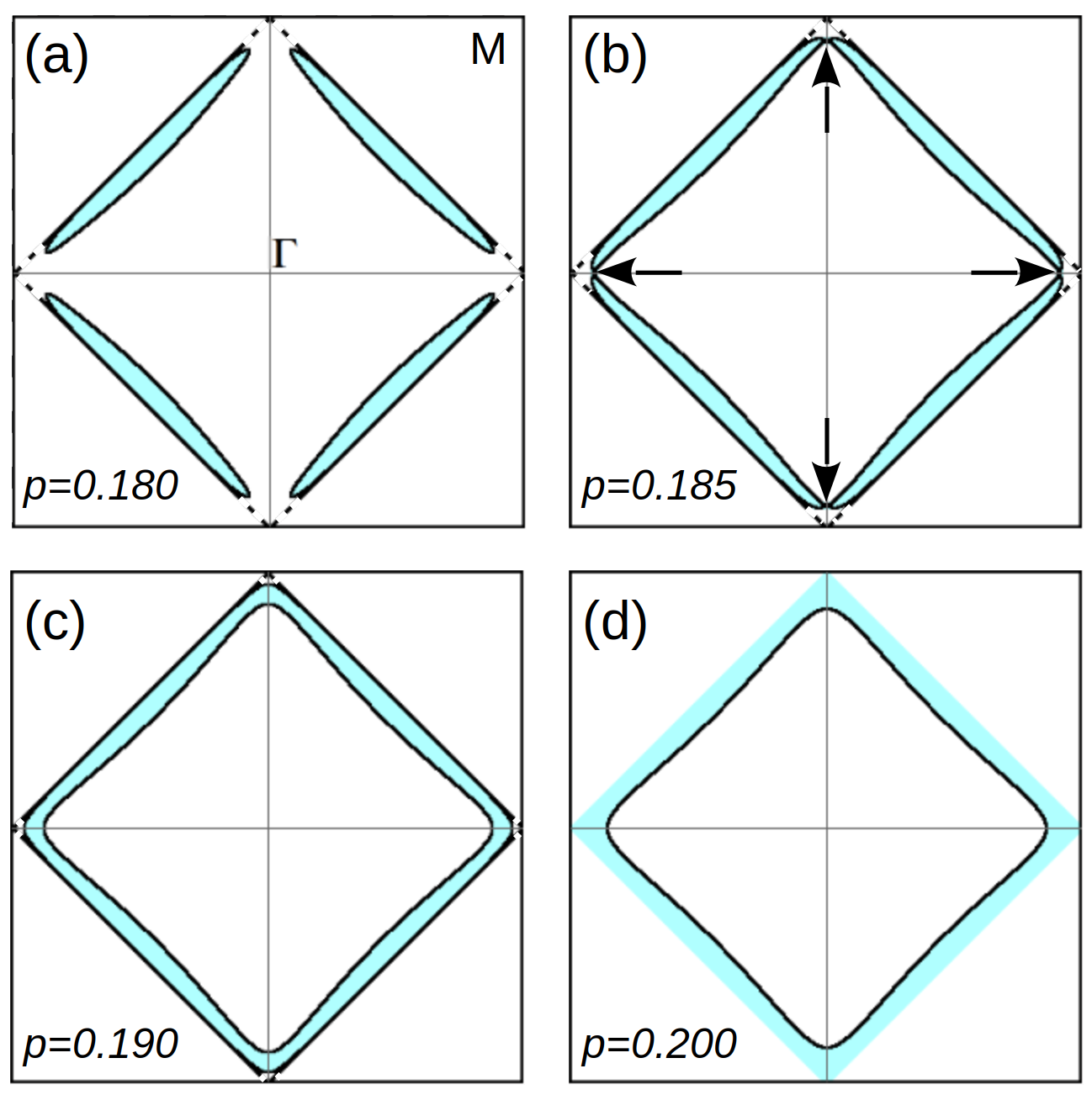}
\caption{(Color Online) Fermi surface evolution with doping near the pseudogap end-point. Hole occupation is
indicated by blue shade. (a) Singular self energy splits a simply connected Fermi surface into hole pockets. (b)
The hole pockets enlarge with doping, and they touch at exceptional van Hove points (indicated by arrows), located
on high symmetry lines, but not on high symmetry points. (c) Further doping forms annular rings of holes.
(d) When the pseudogap vanishes a closed electron-like weakly interacting Fermi surface is recovered.}
\label{fig1}
\end{center}
\end{figure}

Our main observation is the following. Consider a one-band system whose Fermi surface is simply connected
in the weakly interacting limit. Then, the saddle points are necessarily located at high symmetry
points in the Brillouin zone where the Fermi surface can open or close. A typical example is the $(\pm\pi, 0)$ and
$(0, \pm\pi)$ points for a square lattice when the band extremum is at $(0, 0)$ or $(\pi, \pi)$.
This situation is to be contrasted with the case where the interaction is strong enough to induce
self-energy corrections that are singular, such as in a pseudogap phase.
Then, as shown in Fig.~\ref{fig1}(a)-(c),
the self energy splits the simply connected Fermi surface into a multiply
connected surface (i.e, Fermi pockets, or annular Fermi surfaces),
and the saddle point is a result of the touching of two pieces of that surface. In this case
the saddle points are not located on high symmetry points, but they lie on high
symmetry lines, see arrows in  Fig.~\ref{fig1}(b), which has important consequences.
We describe the resulting interaction driven van Hove singularity as being ``exceptional'',
to distinguish them
from ordinary van Hove singularities that are obtained in the weakly interacting limit.

\emph{Model.}---
The canonical system to illustrate the physics of exceptional van Hove singularities are certain
underdoped cuprates in the pseudogap state. Motivated by the
Yang-Rice-Zhang (YRZ) model~\cite{yang06}, we describe it
by a single band of electrons whose Green's function is given by
\beq
\label{eq:G-fn1}
G_{\bk}(i\om_n)^{-1} = i\om_n  - \ep_{\bk} - P_{\bk}^2/(i\om_n  - \xi_{\bk}).
\eeq
This type of model has been justified through
phenomenological~\cite{yang06,norman98,norman07,bascones08,storey16} as well as
numerical~\cite{kyung06,stanescu06,civelli09,liebsch09,sakai09,sakai16,wu18,braganca18,sordi10,sordi11}
cluster dynamical mean field studies of the strong coupling Hubbard model.
An extensive comparison with experiments using the YRZ model has also
been reported in Ref.~\cite{rice12}.
Here, $\ep_{\bk} = -2\tilde{t} (\cos k_x + \cos k_y) - 4t^{\prm} \cos k_x \cos k_y - \mu$ is the
electron dispersion, $\tilde{t} (p) = t[1 - 4(0.2-p)]$ is a hopping parameter
modified by the interaction, $t =1$,  $t^{\prime} = -0.15$, $p$ is the hole
doping, and $\mu$ is the chemical potential.
We take $\xi_{\bk} = 2\tilde{t} (\cos k_x + \cos k_y)$, where the equation $\xi_{\bk} = -\omega$
defines the points on the Brillouin zone where the electron's spectral weight is suppressed due to the
pseudogap. We model the pseudogap by
$P_{\bk}(p) = \theta(p^{\ast} - p) P_0 (1 - p/p^{\ast}) (\cos k_x - \cos k_y)$, where
$\theta(x)$ is the Heaviside step function,
$P_0 =0.4$ is the pseudogap energy at half filling ($p=0$), and which decreases linearly
with hole doping, and terminates at $p^{\ast}=0.2$. All energy scales are in unit of $t$, which we take to
be about 300 meV~\cite{andersen95} for later estimates.

Superficially, Eq.~\eqref{eq:G-fn1} is reminiscent of two hybridizing bands, namely the physical electrons with
dispersion $\ep_{\bk}$ and pseudofermions with dispersion $\xi_{\bk}$. Thus, it can be written as
\beq
\label{eq:G-fn2}
G_{\bk} (i\om_n) = A_{1\bk}/( i\om_n - \om_{1 \bk}) + A_{2\bk}/( i\om_n - \om_{2 \bk}),
\eeq
where
$\om_{1 \bk, 2\bk} = [\ep_{\bk} + \xi_{\bk} \pm \sqrt{(\ep_{\bk} - \xi_{\bk})^2 + 4 P_{\bk}^2}]/2$.
The weight factors $A_{1\bk} = (\om_{1\bk} - \xi_{\bk})/(\om_{1\bk} - \om_{2\bk})$, and
$A_{2\bk} = (\xi_{\bk}- \om_{2\bk})/(\om_{1\bk} - \om_{2\bk})$.
For the doping range studied here only the lower band
$\om_{2\bk}$ contributes to the Fermi surface in the form of hole pockets that evolves with doping,
see Fig.~\ref{fig1} and Fig.~S4 in the Supplementary Information (SI)~\cite{suppl}.
\begin{figure}[t]
\begin{center}
\includegraphics[width=0.4\textwidth]{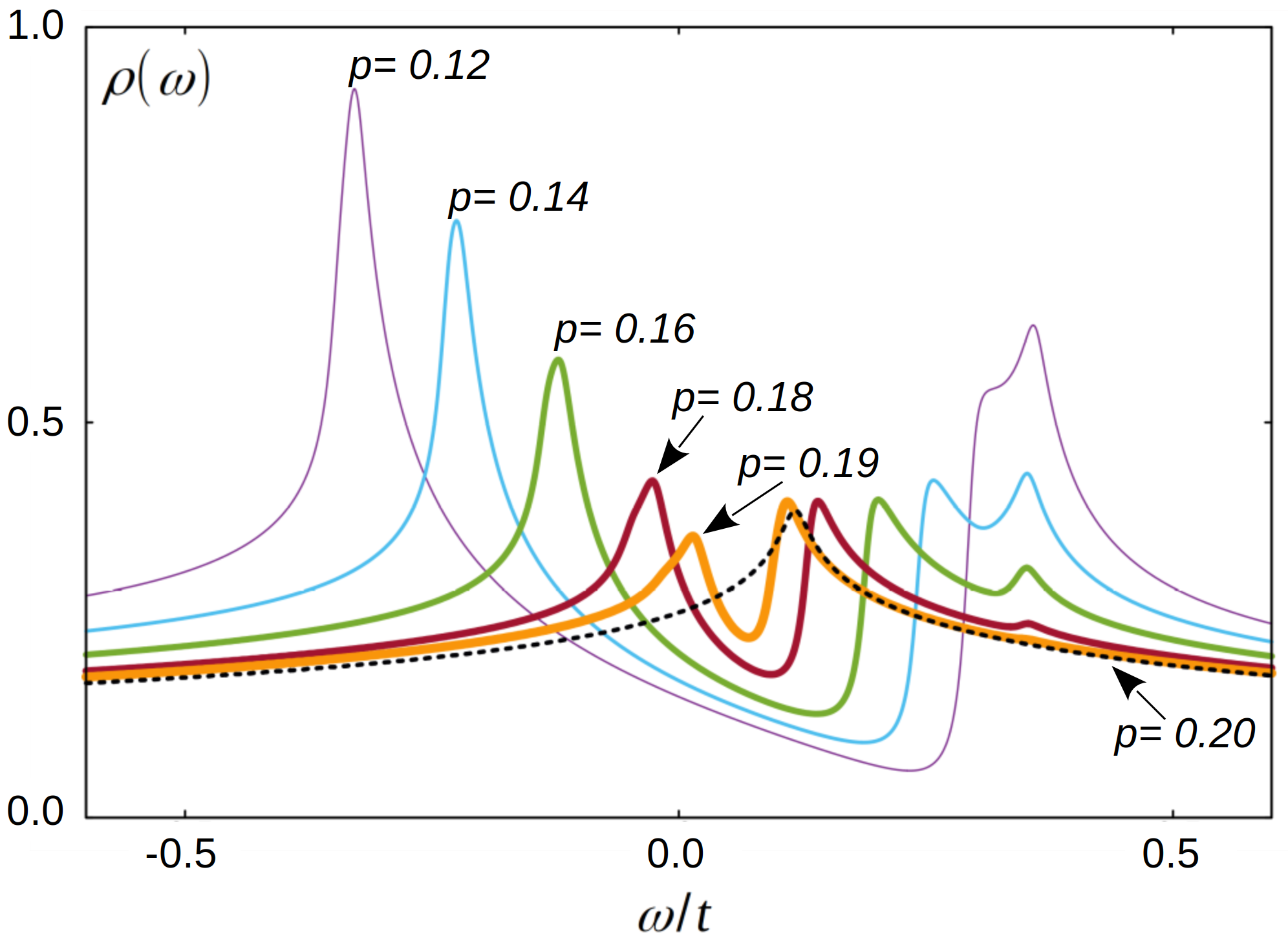}
\caption{(Color Online) Density of states $\rho(\omega)$ for various doping.
The exceptional van Hove singularity manifests
as a peak which is at negative energies $\omega$ for low doping. The peak height diminishes as the
pseudogap potential decreases with doping. The peak crosses $\omega =0$
at $p_{ev} \approx$ 0.185.}
\label{fig2}
\end{center}
\end{figure}

\emph{Exceptional van Hove singularity.}---
As shown in Fig.~\ref{fig1}, with increasing doping the hole pockets grow
and eventually, at a doping $p_{ev} \approx 0.185$, the pockets touch
at the van Hove points $(0, \pm k_{ev})$ and $(\pm k_{ev}, 0)$, where $k_{ev} \neq \pi$,
see arrows in Fig.~\ref{fig1}(b).
The resulting Lifshitz transition describes hole pockets merging to form hole rings,
see  Fig.~\ref{fig1}(c).

In the vicinity of such saddle points, say, the one at $(0, k_{ev})$, the dispersion
can be expressed as $\om_{2\bk} \approx \alpha k_x^2 - \beta k_y^2 - \gamma k_y k_x^2$,
where $(\alpha, \beta, \gamma)$ are parameters with dimension of energy, and $\gamma \neq 0$
indicates that the saddle point is not on a high symmetry location.
The peak in the density of states
$\rho(\omega) \equiv - (1/\pi) \sum_{\bk} {\rm Im} G_{\bk} (\om + i \Gamma)$
near the singularity is given by $\rho(\om) \approx 4 \rho_{sp}(\om)$, where
\begin{align}
\label{eq:DOS}
\rho_{sp}(\om) &= \frac{1}{2 \pi^2 \sqrt{\alpha \beta}} \left[ {\rm Re}
\left[ \frac{1}{(1+u)^{1/4}} K(r_1) \right]
\right. \nonumber \\
& \left.
- {\rm Im} \left[ \frac{1}{(1+u)^{1/4}} K(r_2) \right] \right].
\end{align}
Here, $u = (\om + i \Gamma)/E_0$, $E_0 = \alpha^2 \beta/\gamma^2$,
$r_{1,2}^2 = [1 \pm 1/(1+u)^{1/2}]/2$, and $K(r) \equiv \int_0^{\pi/2} d \theta/\sqrt{1- r^2 \sin^2 \theta}$
is the complete elliptic integral of the first kind, and
$\Gamma= 0.01 t$ is a frequency independent inverse lifetime.

In Fig.~\ref{fig2} we show the evolution of the peak in the density of states with doping. As
$p \rightarrow p_{ev}$, the peak position moves from negative energies $\omega < 0$, and
approaches the chemical potential $\omega =0$. However, for the cuprates, increasing doping
also implies a reduction in the pseudogap strength $P_{\bk}(p)$. Therefore, the strength
of the singularity decreases upon approaching the Lifshitz transition. This is seen as diminishing peak height
of $\rho(\omega)$ with doping in Fig~\ref{fig2}.

\begin{figure}[!!tb]
\begin{center}
\includegraphics[width=0.48\textwidth]{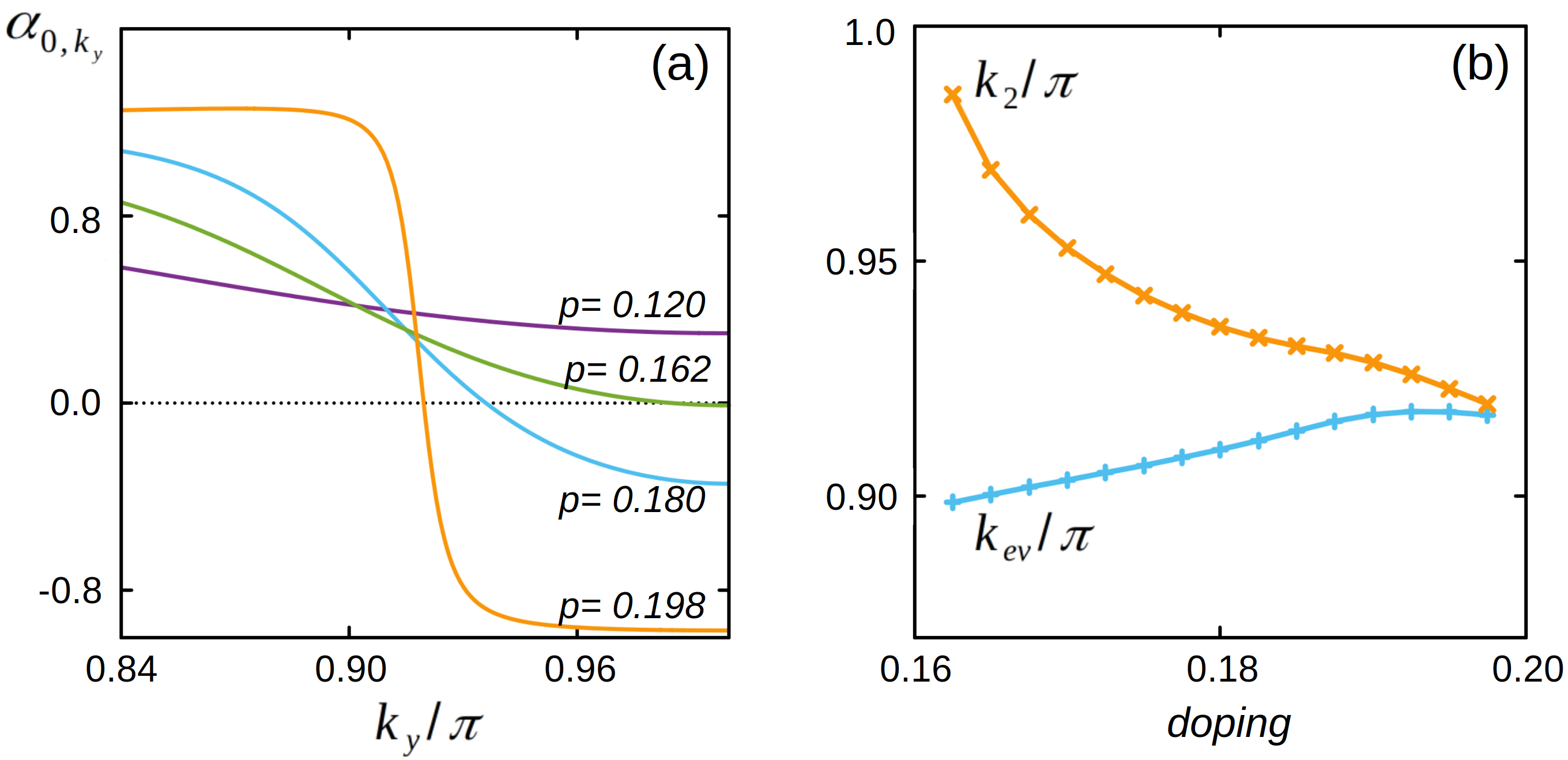}
\caption{(Color Online) (a) Plot of $\alpha_{\bk} \equiv \partial^2 \om_{2 \bk}/\partial k_x^2$
along the $(0, 0) - (0, \pi)$ direction for various doping. For $p > 0.162$ in the current model, the
curvature $\alpha_{\bk}$ vanishes at $(0, k_2)$ and symmetry equivalent points.
(b) Variation of $k_2$ and $k_{ev}$ with doping. Near the pseudogap
end-point the exceptional van Hove singularity is close to a second order one, and
the pre-factor of the log in the density of states is large.}
\label{fig3}
\end{center}
\end{figure}

\emph{Proximity to second order van Hove singularity.}---
This is a consequence of $\gamma \neq 0$.
In Fig.~\ref{fig3}(a) we plot the curvature $\alpha_{\bk} \equiv \partial^2 \om_{2 \bk}/\partial k_x^2$
along the $(0, 0) - (0, \pi)$ direction for various doping. We notice that, when the pseudogap is sufficiently
small ($p\ge 0.16$),
$\alpha_{(0, k_y)}$ has positive values at $k_y \sim 0$, and it has negative values at $k_y \sim \pi$,
implying it goes through zero at $ k_y = k_2 \sim \alpha/\gamma$.

If $\alpha_{\bk} = \bv_{\bk} = 0$ is simultaneously satisfied,
the system has a second order van Hove singularity~\cite{tamai08,shtyk17,efremov19,yuan19,hsu21,guerci22}.
The density of states has a power law singularity, obtained by taking $\alpha \rightarrow 0$ in
Eq.(\ref{eq:DOS}), instead of the usual log singularity.
In our case these two points are located close by
on the same high symmetry lines, i.e. $k_{ev} \sim k_2$, implying that the system is close to the second order singularity, and therefore, the pre-factor of the log is large.
Indeed, we find that as $p \rightarrow p^{\ast}$, the wavevectors $k_{ev}$ and $k_2$ come closer,
as shown in Fig.~\ref{fig3}(b). As shown in Figs.~S1 and S2 of
the SI~\cite{suppl}, the conversion to second order singularity
can be readily achieved by varying a third nearest neighbor hopping parameter which,
a priori, is feasible in cold atom systems.

Note, for a single band system with a simply connected Fermi surface, no such proximity to a
second order van Hove singularity is expected. Thus, in such systems, this proximity distinguishes an
interaction induced exceptional van Hove singularity from a weakly interacting ordinary one. In multiband
systems, however, higher order van Hove singularities can arise from
noninteracting physics alone~\cite{tamai08,shtyk17,efremov19,yuan19,hsu21,guerci22}.

\emph{Exceptional van Hove singularity near pseudogap end-point.}---
In the rest of this work we examine the relevance of exceptional van Hove singularities for the
cuprates in the vicinity of the hole doping $p^{\ast}$ where the pseudogap terminates.

First, we explain why, for a large class of cuprates, the chemical potential necessarily
crosses the exceptional van Hove singularity as the pseudogap end-point is approached from the
underdoped side. As a function of doping (or equivalently, the chemical potential) the evolution
of the Fermi surface can be tracked by solving the equation
Re[$G_{\bk}(\om =0)^{-1}$]$=0$ .
In the absence of any pseudogap term, the self energy corrections are
analytic, and therefore the Fermi surface is bound to cross the $(0, \pi)$ and $(\pi, 0)$ points at a
doping $p_0$ where the ordinary (or the weakly interacting) van Hove singularity
crosses the chemical potential, thereby
transforming from a open to a closed Fermi surface of electrons. In the presence of the pseudogap,
however, at $\om =0$ the Fermi surface cannot cross these points since the pseudogap term is a divergent
repulsive potential on the manifold $\xi_{\bk}=0$ where these points lie. Generally, there are three possible
ways in which the Fermi surface evolution can respond to this divergence.

First, the possibility that the
pseudogap terminates at a doping exactly where the chemical potential crosses the
ordinary van Hove singularity, i.e.,
$p^{\ast} = p_0$. In this case the open Fermi surface can close smoothly by crossing the
 $(0, \pi)$ and $(\pi, 0)$ points. However, this will occur only if the system is fine-tuned. Since such a coincidence is unlikely in general, we do not discuss it further.

The second possibility is that the pseudogap terminates before the ordinary van Hove singularity is reached,
i.e. $p^{\ast} < p_0$. In this case, for the doping range $p_0 > p > p^{\ast}$,
the pseudogap term is zero, the self-energy
is analytic, and therefore,  the open Fermi surface can close smoothly by crossing the
$(0, \pi)$ and $(\pi, 0)$ points.  A recent study has shown that this is indeed the
case for cuprates for which $|t^{\prime}/t|$ is sufficiently large, such as
YBa$_2$Cu$_3$O$_{7-\delta}$ (YBCO),
Tl$_2$Ba$_2$CuO$_{6+\delta}$, and HgBa$_2$CuO$_{4+\delta}$~\cite{wu18}.

The third possibility, namely  $p^{\ast} > p_0$ triggers the exceptional van Hove singularity studied here.
It is relevant for cuprates for which $|t^{\prime}/t|$ is sufficiently small, such as
Bi$_2$Sr$_2$CaCu$_2$O$_{8+\delta}$, (Bi, Pb)$_2$(Sr, La)$_2$CuO$_{6 +\delta}$,
La$_{2-x}$Sr$_x$CuO$_4$ (LSCO), and (Nd, Eu)-LSCO~\cite{wu18}.
In this case, the hole pockets grow with increasing
hole doping, but simultaneously the Fermi surface avoids the  $(0, \pi)$ and $(\pi, 0)$ points, since
$P_{\bk} \neq 0$. In such a
situation there is invariably
a doping $p_{ev}$, with $p^{\ast} > p_{ev} > p_0$, for which the chemical potential
crosses the exceptional van Hove singularity and the hole pockets touch at points which lie on the high
symmetry lines such as $(0,0) - (0, \pi)$, see Fig.~\ref{fig1}(b). Beyond the Lifshitz transition, for
$p^{\ast} > p > p_{ev}$, the hole pockets merge into a ring of holes with annular Fermi surfaces,
see Fig.~\ref{fig1}(c). Then, as $p \rightarrow p^{\ast}$, the
inner Fermi surface of the hole-ring merges with $\ep_{\bk} =0$ and the outer Fermi surface of the
hole-ring merges with the line of Luttinger zeroes $\xi_{\bk} =0$, see Fig.~\ref{fig1}(d).

Therefore, we predict that, in the narrow doping range $p_{ev} < p < p^{\ast}$ separating the
Lifshitz transition from the pseudogap end-point, the holes form an annular ring bounded by two
Fermi surfaces, see Fig.~\ref{fig1}(c), with the physical electron weight mostly on the inner
Fermi surface. From an angle resolved photoemission perspective this is a state with a closed Fermi surface,
but where the $(\pi, 0)$ point is still pseudogapped. Note, this narrow doping range may be difficult
to resolve in an experiment. In that case it will appear that the Lifshitz transition and the
closing of the pseudogap occur at the
same doping, as reported in the recent literature~\cite{Benhabib15,loret17,leyraud17,wu18,braganca18}.

\emph{Experimental signatures.}---
Recently, signatures of unusual thermodynamics have been reported for several cuprates close to the
pseudogap endpoint $p^{\ast}$. Thus, the specific heat coefficients $\gamma(T)$ of LSCO,
(Nd, Eu)-LSCO, Ca doped YBCO and Bi$_{2+y}$Sr$_{2-x-y}$La$_x$CuO$_{6+\delta}$
show logarithmic $T$-dependence~\cite{Klein19,girod21}, while the nematic susceptibility
$\chi_{B_{1g}}$ of Bi$_2$Sr$_2$CaCu$_2$O$_{8+\delta}$ increases considerably in this doping
range~\cite{Auvray19,Benhabib15}. Both these observations can be potentially
explained by the presence of a nematic
quantum critical point (QCP) around $p^{\ast}$, but no such QCP has been
identified until now. Moreover, even though an ordinary van Hove singularity can lead to
$\gamma(T) \sim \log T$, the observed anomaly was deemed too sharp for such an
explanation~\cite{Klein19,girod21}.
As we show below, the above experimental puzzles can be resolved if we assume the presence of
an exceptional van Hove singularity, i.e., these systems follow the third possibility discussed above.

\begin{figure}[t!!]
\begin{center}
\includegraphics[width=0.4\textwidth]{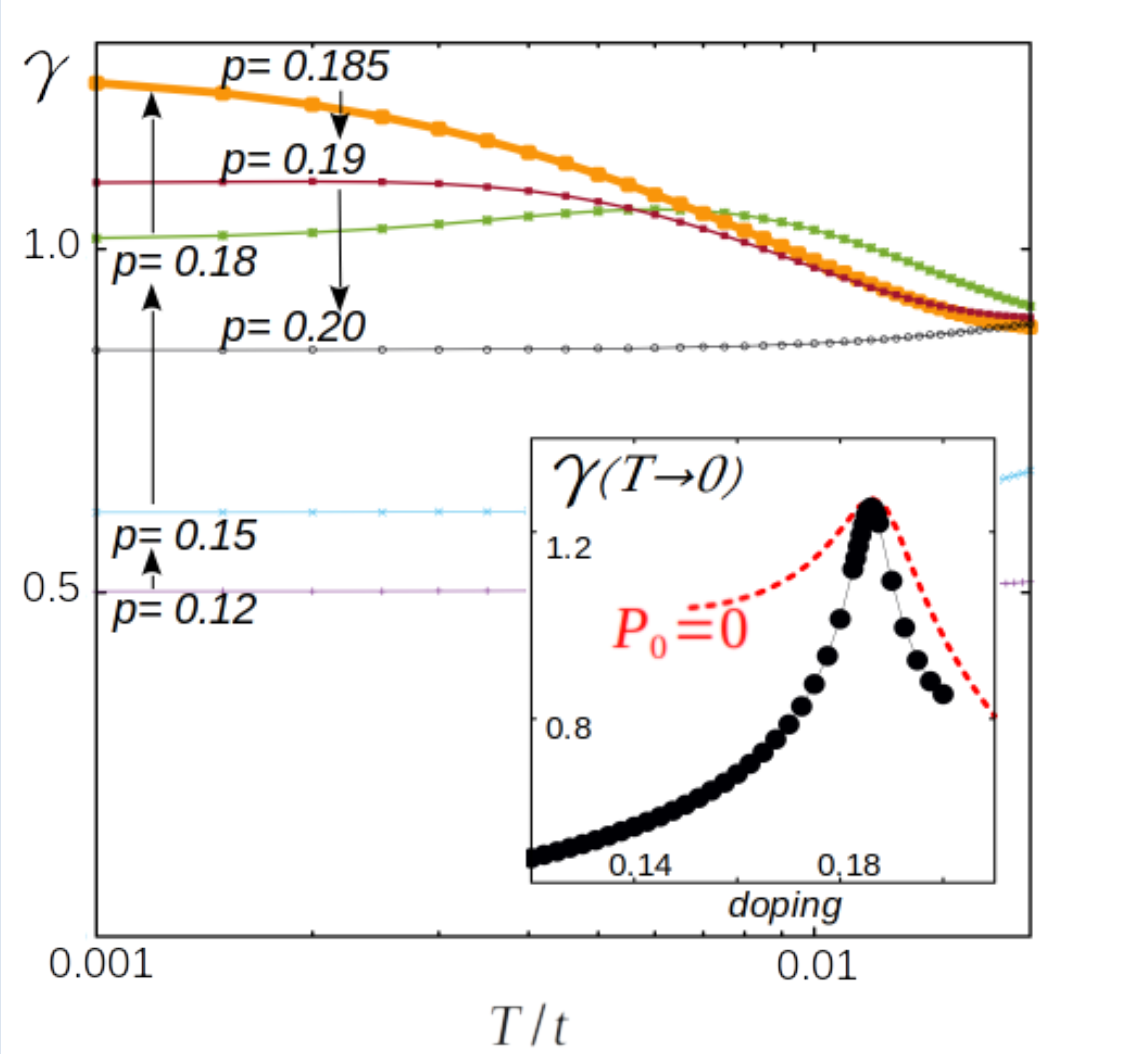}
\caption{(Color Online) Specific heat coefficient $\gamma$ as a function of temperature and doping.
Arrows indicate that $\gamma$ increases as $p \rightarrow p_{ev}$, where the exceptional van Hove
singularity is on the chemical
potential.  At $p = p_{ev}$, the saturation of $\gamma(T)$ with temperature is pushed to lower $T$.
The inset shows the doping dependence of $\gamma (T \rightarrow 0)$ (black dots)
with a sharp peak at $p = p_{ev}$. Red dotted line is the calculation for an ordinary van Hove
singularity obtained for pseudogap $P_{\bk}=0$, and shifted horizontally and vertically for better
comparison. The peak is sharper and more pronounced for an exceptional van Hove singularity.}
\label{fig4}
\end{center}
\end{figure}

We compute the specific heat coefficient $\gamma (T) = C(T)/T$ using
\[
C(T) = \int_{-\infty}^{\infty} d \om \frac{ \om^2/T^2}{\cosh^2[\om/(2T)]} \rho(\om),
\]
and we show its $(p, T)$ evolution in
Fig.~\ref{fig4}. As $p \rightarrow p_{ev}$, there is a distinct upturn in the
$T$-dependence of the specific heat coefficient, reminiscent of what has been reported for
in Refs.~\cite{Klein19,girod21} with $\gamma(T) \sim \log T$. In our computation
there is indeed such a logarithmic component.
However, the overall $T$- dependence in our theory is way more complex, just as is the
frequency dependence of the density of states in Eq.~\eqref{eq:DOS}.

The $p$ dependence of $\gamma(T \rightarrow 0, p)$ is also
similar to what has been reported in Refs.~\cite{Klein19,girod21}. As shown in
Fig~\ref{fig4} inset, it has a sharp maxima around $p=p_{ev}$, which is close to,
but not coincident with $p^{\ast}$.
This sharp feature can be contrasted with the weak coupling case by setting the pseudogap term
$P_{\bk} =0$. As shown by the red dotted line in Fig~\ref{fig4} inset, the peak is broader and
far less pronounced for an ordinary
van Hove singularity compared to an exceptional one.
This is because the latter is close to a second order
van Hove instability. Also, the bands are flatter than usual for the exceptional
case since the velocity vanishes at two
proximate points $(0, k_{ev})$ and $(0, \pi)$, instead of only at $(0, \pi)$.

\begin{figure}[t!!]
\begin{center}
\includegraphics[width=0.4\textwidth]{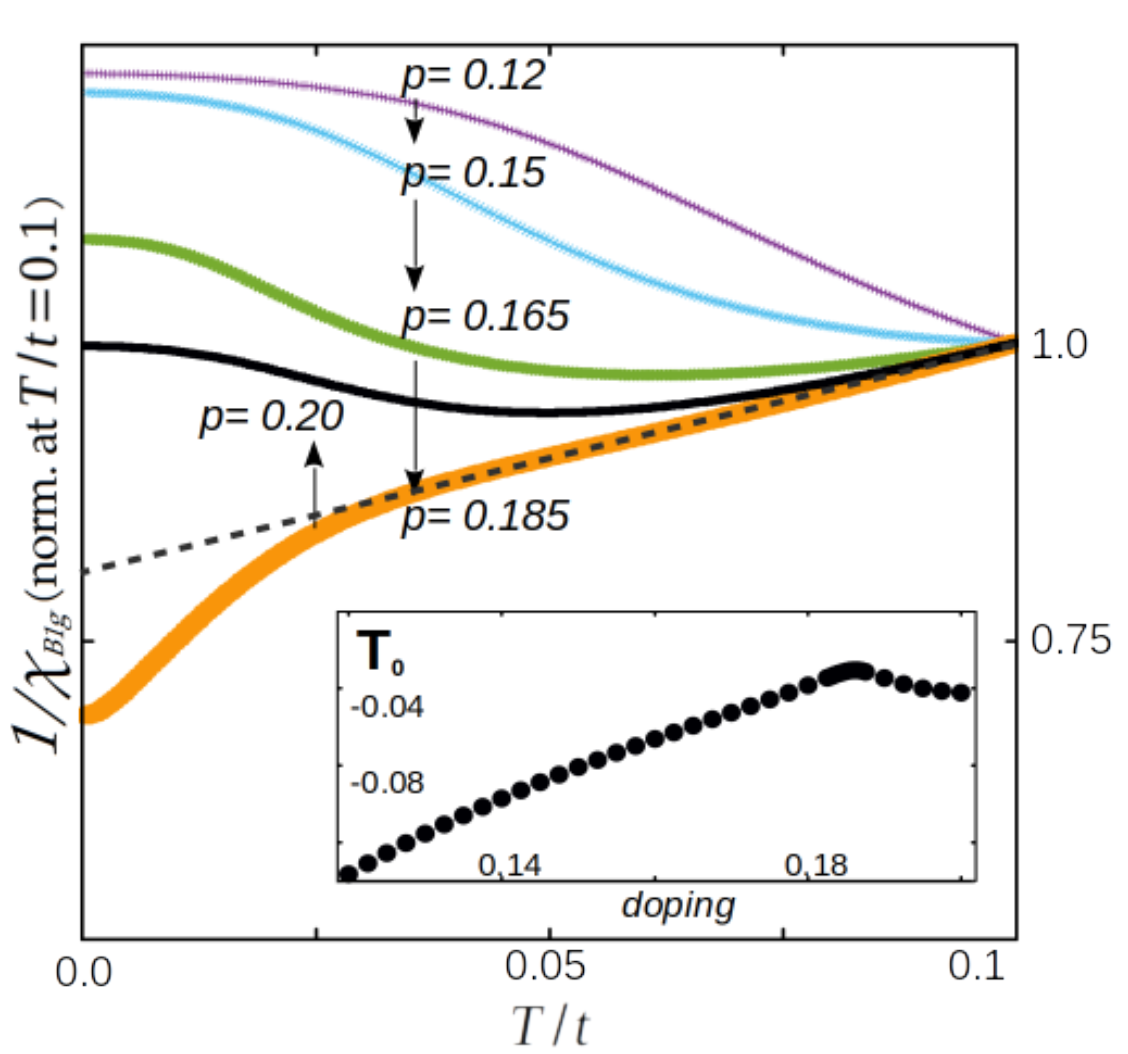}
\caption{(Color Online) Temperature and doping dependence of the inverse nematic susceptibility
$\chi_{B_{1g}}^{-1}$ in the
$(x^2-y^2)$ channel. The $T$-dependence enhances considerably around $p_{ev}$,
with Curie-Weiss $1/T$ scaling above a cutoff temperature (black dash line).
The inset shows the doping dependence of
$T_0 \equiv \chi_{\chi_{B_{1g}}}^{-1} (T \rightarrow 0)$ with a peak near
$p_{ev}$.
}
\label{fig5}
\end{center}
\end{figure}

Next, we study the nematic susceptibility in the $B_{1g}$ or the
$(x^2-y^2)$ symmetry channel, which is given by
\[
\chi_{B_{1g}} = - {\rm Im} \left[ \sum_{\bk} h_{\bk, B_{1g}}^2 \int_{-\infty}^{\infty} \frac{d \om}{2 \pi}
 \tanh [\frac{\om}{2T}] G_{\bk}(\om + i\Gamma)^2 \right],
\]
with $h_{\bk, B_{1g}} = \partial^2 \ep_{\bk}/\partial k_x^2 - \partial^2 \ep_{\bk}/\partial k_y^2$.
In Fig.~\ref{fig5} we present the $(p, T)$
dependencies of $\chi_{B_{1g}}^{-1}$. It has weak $T$-dependence
away from the critical doping $p_{ev}$, while at $p_{ev}$  it decreases
considerably with lowering temperature (equivalently,  $\chi_{B_{1g}}(T)$ increases with
lowering $T$), which
can be fitted to a Curie-Weiss behavior above a temperature cutoff.
Taking $t \sim$ 300 meV, the temperature range of the Curie-Weiss behavior in our calculation
is 100 K to 300 K, which matches well with the range seen in the experiments.
Thus, our result is qualitatively
close to what has been reported for Bi$_2$Sr$_2$CaCu$_2$O$_{8+\delta}$
in Fig. 2(b) of Ref.~\cite{Auvray19}. Note, the low temperature
downturn in $\chi_{B_{1g}}^{-1}$ for $p \sim p_{ev}$ seen in Fig.~\ref{fig5}
is a theoretical prediction that can be
checked by performing Raman spectroscopy at lower temperatures. Finally, the inset of Fig.~\ref{fig5}
shows the doping dependence of $T_0 \equiv \chi_{B_{1g}}^{-1}(T=0)$. Once again, the
non-monotonic $p$ dependence, with $T_0$ coming close to zero
(equivalently, large $\chi_{B_{1g}}$) around $p \sim p_{ev}$,  captures what is
reported in Fig.~3 of Ref.~\cite{Auvray19}.

\emph{Conclusion.}---
In summary, motivated by the pseudogap state of the cuprates,
we introduced the concept of a strong interaction driven ``exceptional'' van Hove singularity.
It appears when
the single particle dispersion has a singular correction
that splits an otherwise simply connected Fermi surface into multiply connected pieces.
The exceptional van Hove singularity describes the touching of two pieces of the split Fermi surface.
The associated saddle points of the
renormalized dispersion are located, not on high symmetry points, but on high symmetry lines of the
Brillouin zone.
This feature is proximate to a second order van Hove
singularity. Consequently, the logarithmic
divergence of the density of states is guaranteed to have a large prefactor.
Most importantly, we argued that
several hole doped cuprates necessarily encounter an exceptional van Hove point as they
approach the pseudogap end point, and we showed
that the signatures of the singularity can explain recent experiments~\cite{Klein19,Auvray19,Benhabib15,girod21}.
We expect the electronic dispersion to show features of the exceptional van Hove
which can be detected by
photoemission~\cite{loret18,zhong22} (see discussion associated with Figs.~S3 - S5
of the SI for further details~\cite{suppl}).

Finally, our work can be extended in two new directions in the
future. First, to study interaction driven second van Hove singularities. Second,
to study exceptional van Hove singularities in heavy fermions, where the Kondo coupling
with the localized spins provide a singular correction to the conduction electron dispersion.

We are thankful to M. C. O. Aguiar, Y. Gallais, F. Pi\'{e}chon, A. Sacuto,
and L. Taillefer for insightful discussions. We acknowledge financial support from French
Agence Nationale de la Recherche (ANR) grant ANR-19-CE30-0019-01 (Neptun).

\end{document}


\title{
Supplementary Information for ``Exceptional van Hove Singularities in Pseudogapped Metals''
}
\date{\today}
\author{Indranil Paul$^1$ and Marcello Civelli$^2$}
\affiliation{
$^1$Universit\'{e} Paris Cit\'{e}, CNRS, Laboratoire Mat\'{e}riaux et Ph\'{e}nom\`{e}nes Quantiques,
75205 Paris, France\\
$^2$Universit\'{e} Paris-Saclay, CNRS, Laboratoire de Physique des Solides, 91405, Orsay, France
}
\maketitle
\subsection{Accessing the second order van Hove singularity by tuning a hopping parameter}

The purpose of this section is to demonstrate that the exceptional van Hove singularity can be
converted into a second order singularity by tuning a hopping parameter. Thus,
the logarithmic singularity of  the density of states can be fine tuned to a power-law one.
Since, in the space of tuning parameters, the exceptional van Hove singularity is close to a second
order one, this would imply that, even without fine tuning, the pre-factor of the log in the density of
states is typically large.
\begin{figure}[b]
\begin{center}
\includegraphics[width=0.4\textwidth]{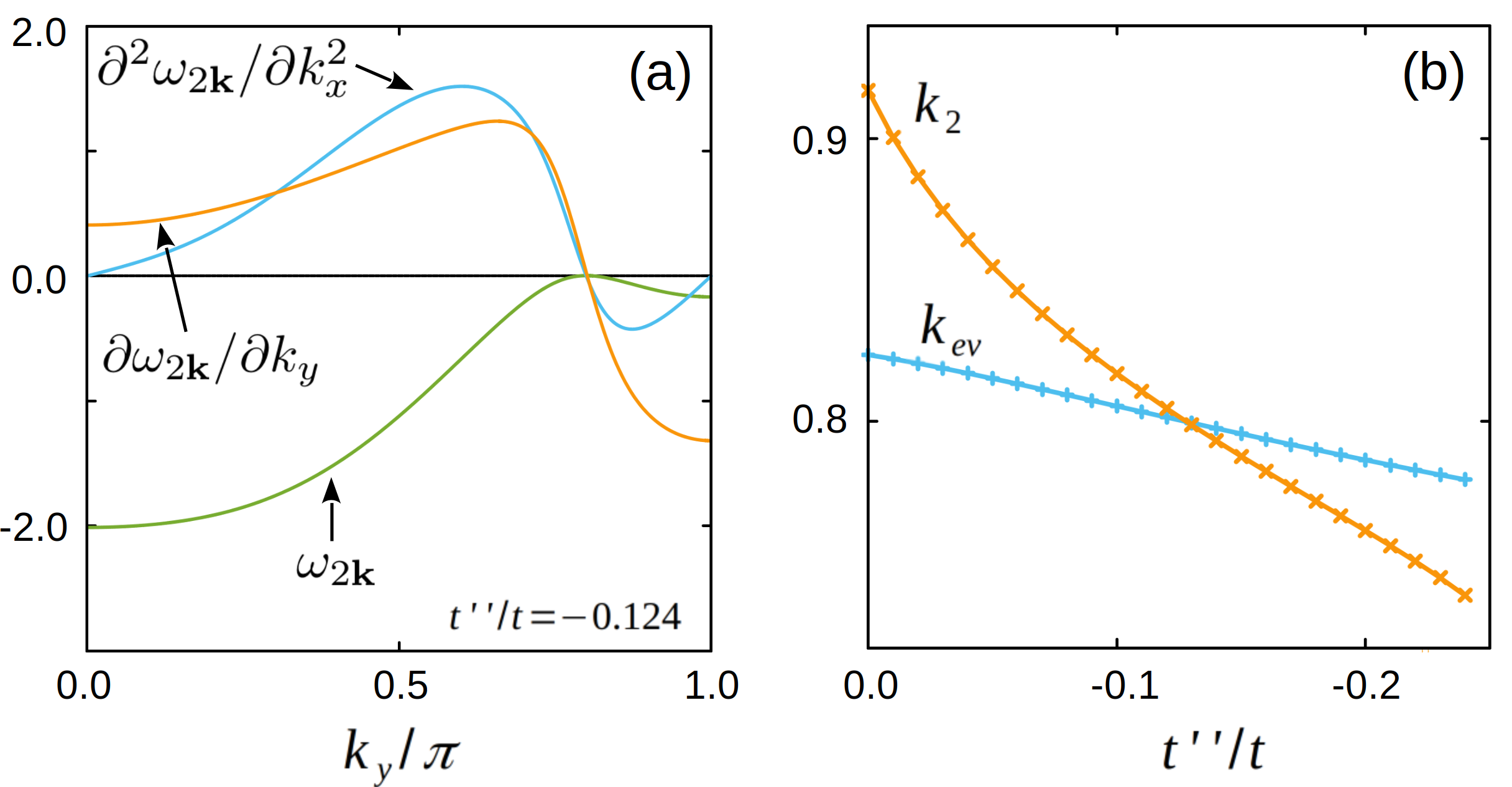}
\caption{(Color Online) (a)  Plot of the dispersion $\omega_{2\bk}$ (green), the velocity
$\partial \omega_{2\bk}/\partial k_y$ (orange) and the curvature
$\partial^2 \omega_{2\bk}/\partial k_x^2$ (blue) along the high symmetry line $(0, 0)$ to
$(0, \pi)$. All the three curves go to zero at the same $\bk$ point, indicating that the system at
a second order van Hove singularity. (b) Variations with third nearest neighbor hopping parameter
$t^{\prime \prime}$ of the points $k_{ev}$ and $k_2$, where the velocity
$\partial \omega_{2\bk}/\partial k_y$ and the curvature $\partial^2 \omega_{2\bk}/\partial k_x^2$
goes to zero, respectively, along the same high symmetry line. Where they coincide the system is fine
tuned to a second order van Hove singularity.
}
\label{figS1}
\end{center}
\end{figure}

\begin{figure}[t]
\begin{center}
\includegraphics[width=0.4\textwidth]{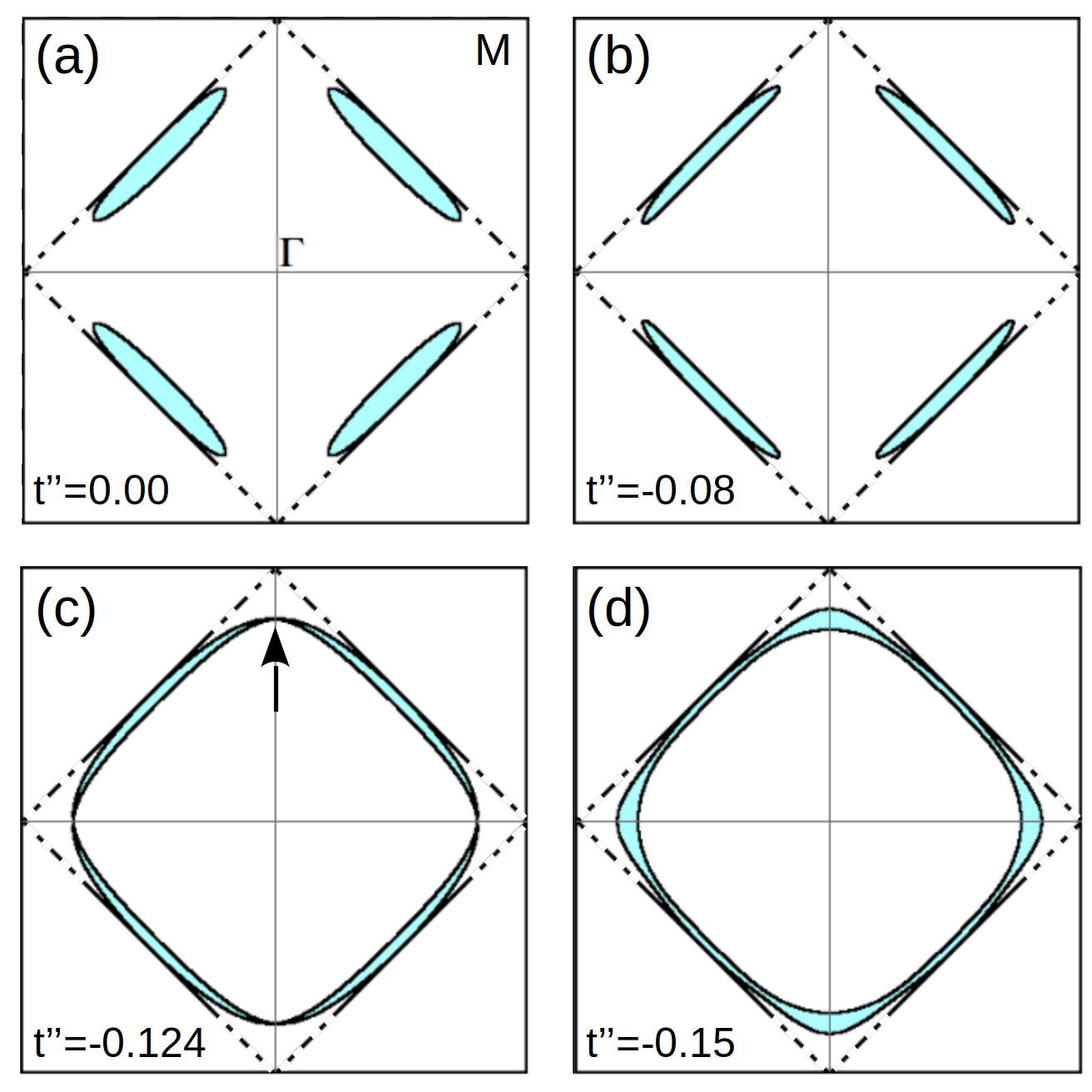}
\caption{(Color Online) (a) - (d) Evolution of Fermi surface with the tuning of the third nearest
neighbor hopping parameter $t^{\prime \prime}$ as the system undergoes a Lifshitz transition
in panel (c), where the van Hove point is marked by an arrow.
The exceptional van Hove singularity is now of second order as demonstrated in Fig.~\ref{figS1}.
}
\label{figS2}
\end{center}
\end{figure}

Accordingly, we consider the same structure of the electronic Green's function as given
by Eq.~(1) of the main text, with the bare dispersion now given by
$
\ep_{\bk} = -2t (\cos k_x + \cos k_y) - 4t^{\prm} \cos k_x \cos k_y
- 2 t^{\prime \prime} (\cos 2k_x + \cos 2k_y) - \mu,
$
$t=1$, $t^{\prime}=-0.15$, $\mu = -0.885$. We also set the pseudogap scale $P_0 =0.2$.
In other words, we turn on a third nearest neighbor hopping $t^{\prime \prime}$,
which is varied in the following.
By tuning this parameter the exceptional van Hove singularity becomes a second order one for
$t^{\prime \prime}/t \approx -0.124$. This is shown in Fig.~\ref{figS1}. In panel (a) we show
the dispersion of the lower band $\omega_{2\bk}$, its velocity $\partial \omega_{2\bk}/\partial k_y$
and its curvature $\partial^2 \omega_{2\bk}/\partial k_x^2$ along the high symmetry line
$(0, 0)$ to
$(0, \pi)$. All the three curves go to zero at the same $\bk$ point, indicating that the system is at
a second order van Hove singularity. The same can be seen in panel (b) where,
as a function of $t^{\prime \prime}$, we track the variation
of the points $k_{ev}$ and $k_2$, where the velocity
$\partial \omega_{2\bk}/\partial k_y$ and the curvature $\partial^2 \omega_{2\bk}/\partial k_x^2$
goes to zero, respectively, along the same high symmetry line. Since the two curves cross,
at the fine tuned value of $t^{\prime \prime}$ the system has a second order van Hove singularity.
Simultaneously, in Fig.~\ref{figS2} we show the evolution of the Fermi surface as the system goes
across this singularity.

\subsection{Detection of Exceptional van Hove Singularity by
  Angle Resolved Photoemission Spectra}

\begin{figure}[t]
\begin{center}
\includegraphics[width=0.4\textwidth]{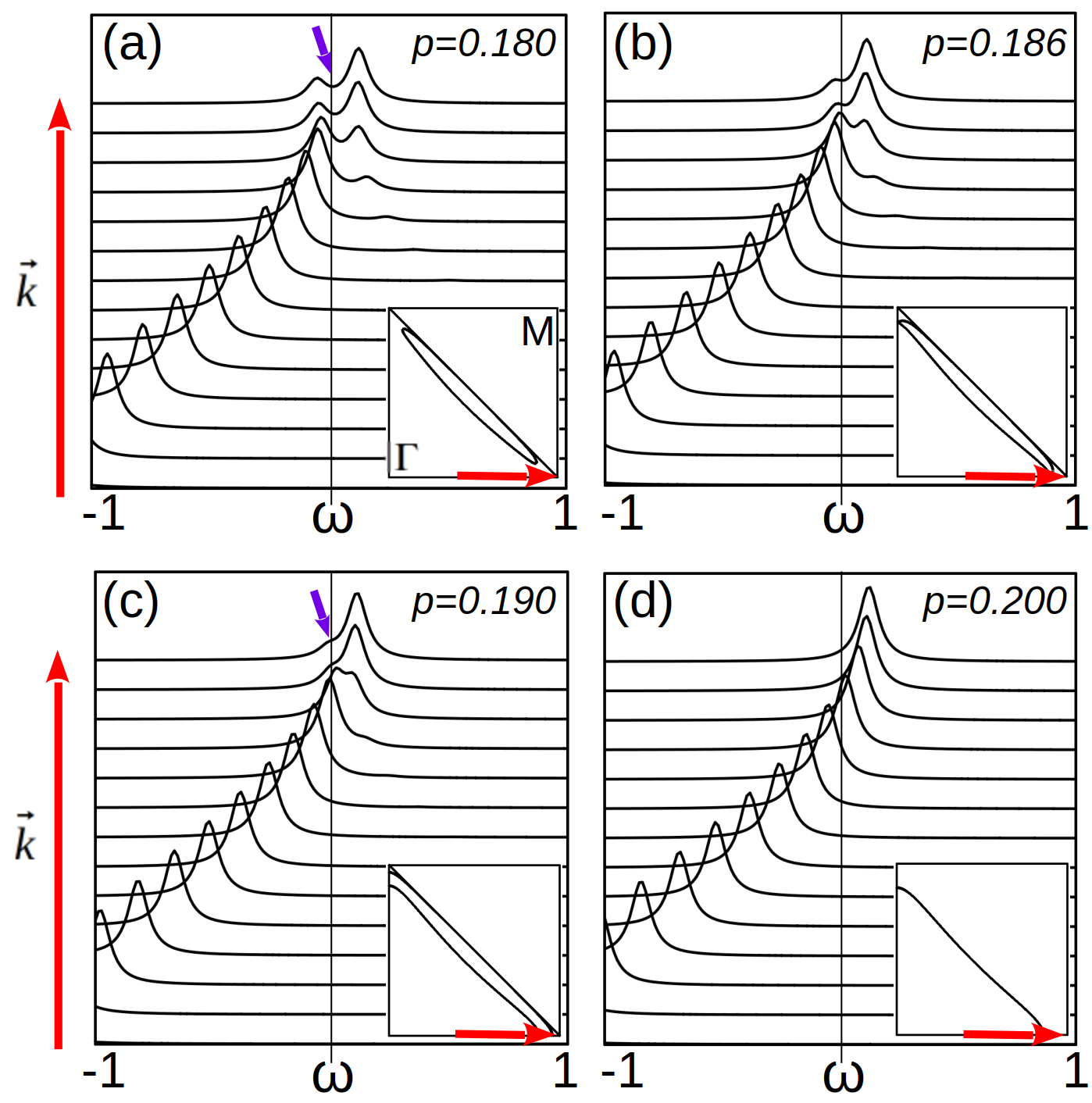}
\caption{(Color Online). Spectral peak dispersion of $A(\mathbf{k},\omega)$ vs $\omega$, accessible
by ARPES,
 plotted along the high symmetry line $(0,0) \to (\pi, 0)$, for various doping $p$
 across the exceptional van Hove singularity. The $x$- and $y$-axes are in unit of $t$ and $1/t$, respectively,
 where $t \sim$ 300 meV.}
\label{figS3}
\end{center}
\end{figure}

We show here that the Lifshitz transition at doping $p = p_{ev}$ associated with the
exceptional van Hove singularity (Fig.~1b of the main text) and the appearance of a ring of annular hole
Fermi surface for $p_{ev}< p< p^{\ast}$ (Fig.~1c of the main text) will have distinct signatures
in the band dispersion which can be detected by
angle resolved photoemission spectroscopy (ARPES). With this motivation, we
display in Fig.~\ref{figS3} the spectral function
(see Eq.~(2) of the main text)
$A(\mathbf{k},\omega)=\, -(1/\pi)\,\text{Im}G_{\bk}(\omega)$,
which is directly related to the photoemission spectra,
as a function of frequency along the high symmetry
$\mathbf{k}$-line $(0,0) \to (\pi, 0)$ of the Brillouin Zone.
The exceptional van Hove point $(k_{ev}, 0)$ lies on the chemical potential at the doping $p_{ev}$
along this path, as shown in Fig.~1b of the main text.
\begin{figure}[b]
\begin{center}
\includegraphics[width=0.4\textwidth]{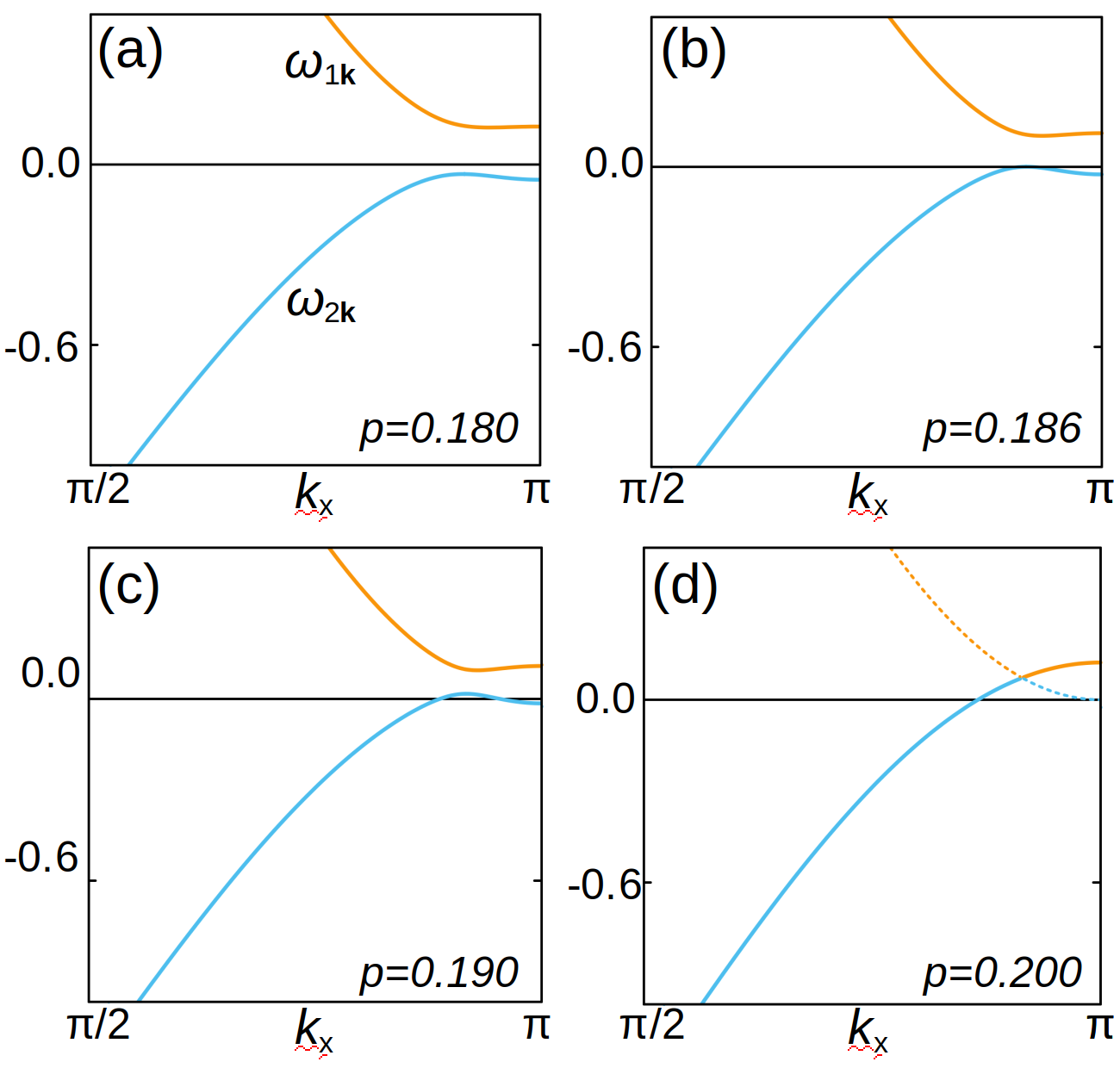}
\caption{(Color Online). Band dispersions
  $\omega_{1\mathbf{k},2\mathbf{k}}$ along the high symmetry
  line $(0,0) \to (\pi, 0)$ for various doping $p$ across the exceptional van Hove singularity.
  Energy is in unit of $t$.}
\label{figS4}
\end{center}
\end{figure}

When the pseudogap is still sufficiently strong, the Fermi surface forms a hole
pocket (see Fig.~1a of the main text). In Fig.~\ref{figS3}.(a) the corresponding spectral function
$A(\mathbf{k},\omega)$ has a peak at $\omega <0$ which,  with increasing $k_x$, disperses towards
the $\omega=0$ line and then bends back, thereby opening a gap at $(\pi, 0)$.
This is the feature that
characterizes the pseudogap in ARPES, and it is a well established result
(see e.g Ref.~\cite{damascelli03}).
However, as the doping $p \rightarrow p_{ev}$, this gap closes in the
$\omega<0 $ side (Fig. \ref{figS3}.(b)).
The peak dispersion tangentially touches the Fermi level at $\omega=0$ for
$k_x= k_{ev}$, and then bends back for increasing $k_x$.
This is the key signature that the exceptional van Hove Lifshitz transition
has occurred. Notice that, from the ARPES
perspective (which mainly has access to the occupied side $\omega<0$ of the spectra),
this point can be interpreted as the doping $p^{\ast}$ where the pseudogap
closes. However, according to our results, this doping, rather, marks $p_{ev}$.
Here the pseudogap is small but
still present, and is visible in the spectra as a small suppression of the spectral weight
just above the Fermi level $\omega>0$.
\begin{figure}[t]
\begin{center}
\includegraphics[width=0.4\textwidth]{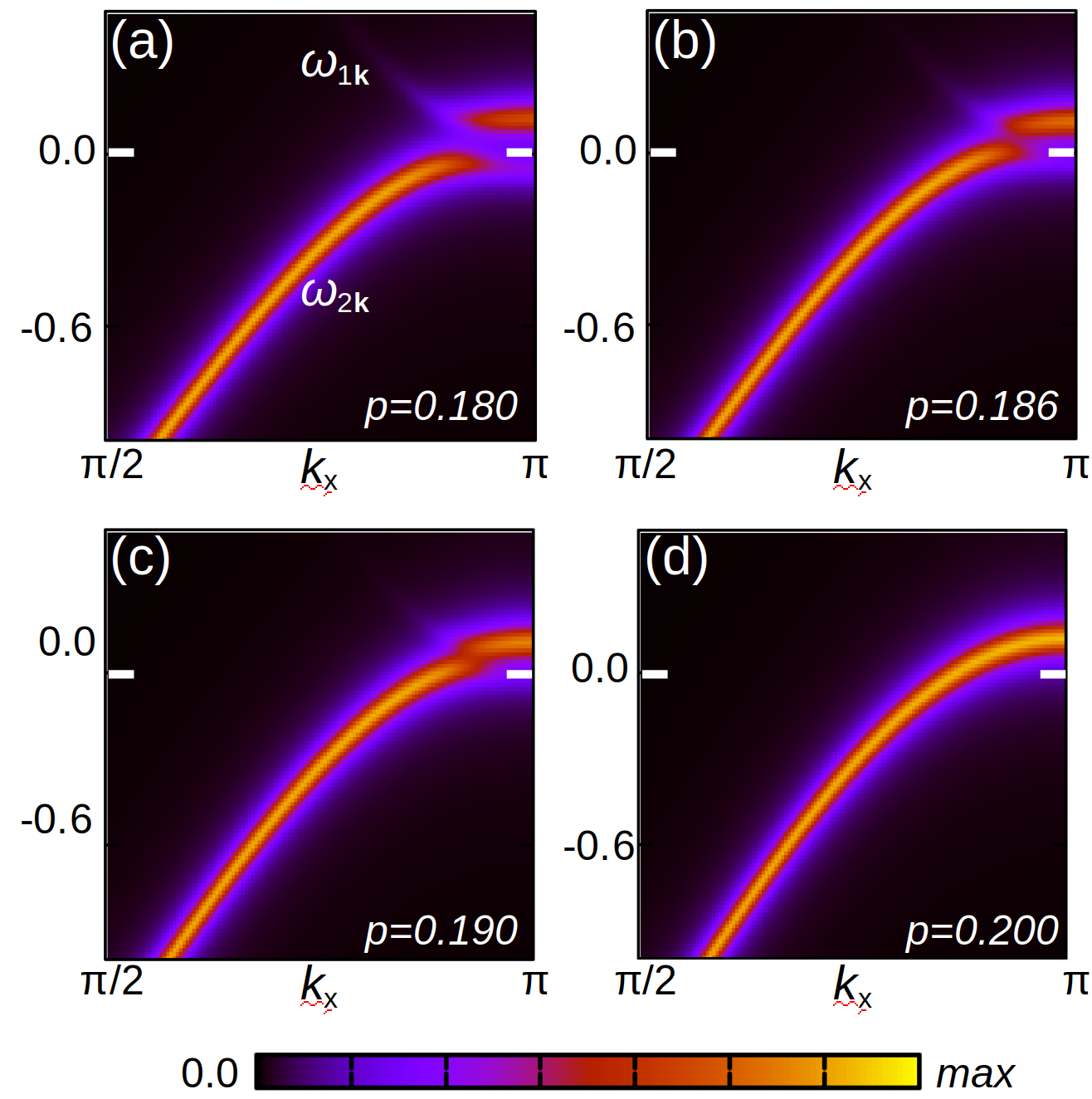}
\caption{(Color Online). Spectral intensity plots $A(\mathbf{k},\omega)$, accessible by ARPES,
plotted along the high symmetry line $(0,0) \to (\pi, 0)$, for various doping $p$
across the exceptional van Hove singularity.
Energy is in unit of $t$.}
\label{figS5}
\end{center}
\end{figure}

By further increasing doping $p_{ev}< p <p^{*}$, a ring Fermi surface
appears (see Fig.~1d of the main text). Therefore we should observe a second
Fermi wavevector on the high symmetry line  $(0,0) \to (\pi, 0)$
and the corresponding quasiparticle peak.
However the outer side of the ring is close to
the large scattering Yan-Zhang-Rice surface $\xi_{\mathbf{k}}=0$,
which strongly suppresses spectral weight.
The quasi-particle peak can be still observed as a
shoulder-like feature, which we indicate
by the small arrow in Fig. \ref{figS3}(c).

Notice that in this ring Fermi surface region $p_{ev}< p <p^{*}$,
a faint trace of the pseudogap spectral weight suppression is
still present in the $\omega>0$ side. This finally disappear
only for or $p\ge p^{\ast}=0.20$, where the more typical
Fermi liquid peak dispersing through $\omega=0$ is recovered (Fig.~\ref{figS3}d).
\begin{figure}[t]
\begin{center}
\includegraphics[width=0.45\textwidth]{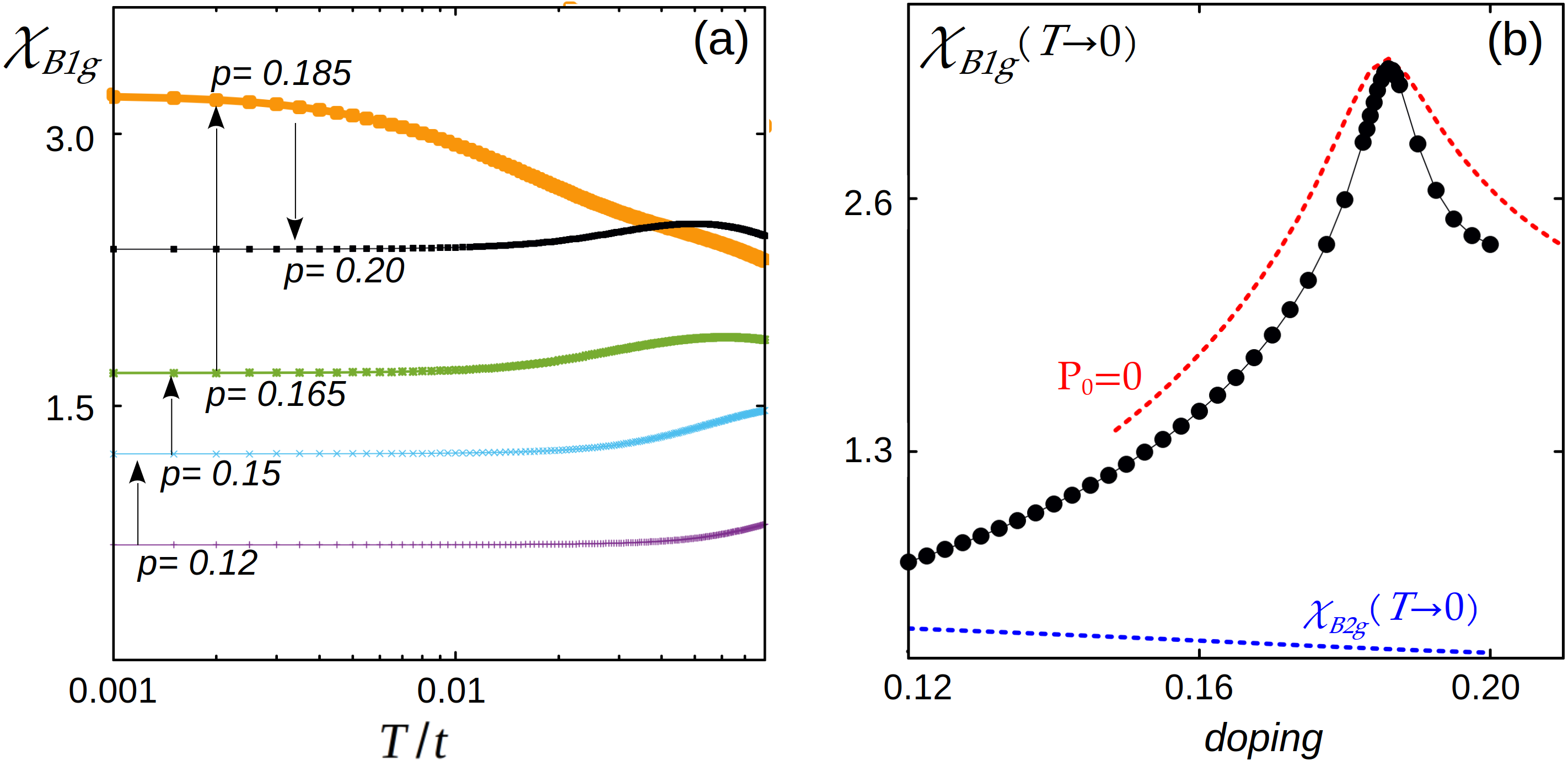}
\caption{(Color Online) (a) Temperature and doping dependence of the nematic susceptibility
$\chi_{B_{1g}}$ in the
$(x^2-y^2)$ channel. (b) Doping dependence of $\chi_{\chi_{B_{1g}}}(T \rightarrow 0)$  showing a peak near
$p \sim p_{ev}$. As in the inset of Fig.~4 of the main text, the peak is sharper and more pronounced
 compared to an ordinary van Hove singularity (red dotted line).
The dashed blue line is the susceptibility $\chi_{B_{2g}}$ in the $xy$ channel,
multiplied by a factor 11.1 to allow for a comparison.
}
\label{figS6}
\end{center}
\end{figure}

These same features, the occurrence of the exception van Hove Lifshitz transition
and the hole ring Fermi surfaces,
could also be observed by looking at the band dispersions
$\omega_{1\mathbf{k},2\mathbf{k}}$ (Fig. \ref{figS4}) and the corresponding spectral
density plots $A(\mathbf{k},\omega)$ (Fig. \ref{figS5}), which are also
accessible by ARPES.
These are plotted again along the high symmetry line in momentum space
$(0,0) \to (\pi, 0)$.
Note, the depth of the hole ring in energy unit, as seen in Fig. \ref{figS4}(c), is of the order of 15 meV,
if we take $t \sim$ 300 meV. Thus, this feature is within the range of resolution of current ARPES experiments.
The $\omega_{1\mathbf{k}}$ in particular shows the contribution from the upper
band, which we did not discuss in the main text,
as it is not relevant for the exceptional
van Hove singularity. It is however crucial to detect the pseudogap
closing at $p^{\ast}$. The observation of  $\omega_{1\mathbf{k}}$ could allow
therefore to definitively distinguish $p^{\ast}$ from the exceptional
van Hove point $p_{ev}$. In practice, the full pseudogap could be determined by
 $\omega_{1\mathbf{k}}- \omega_{2\mathbf{k}}$ in the spectral
density plots of Fig. \ref{figS5}. Such a measure within ARPES
requires to access the unoccupied
states, via thermal excitations for example (as proposed eg in ref.~\cite{yang08}) or
by pump-exciting electrons (as eg in ref. \cite{zonno21}).
\begin{figure}[b]
\begin{center}
\includegraphics[width=0.3\textwidth]{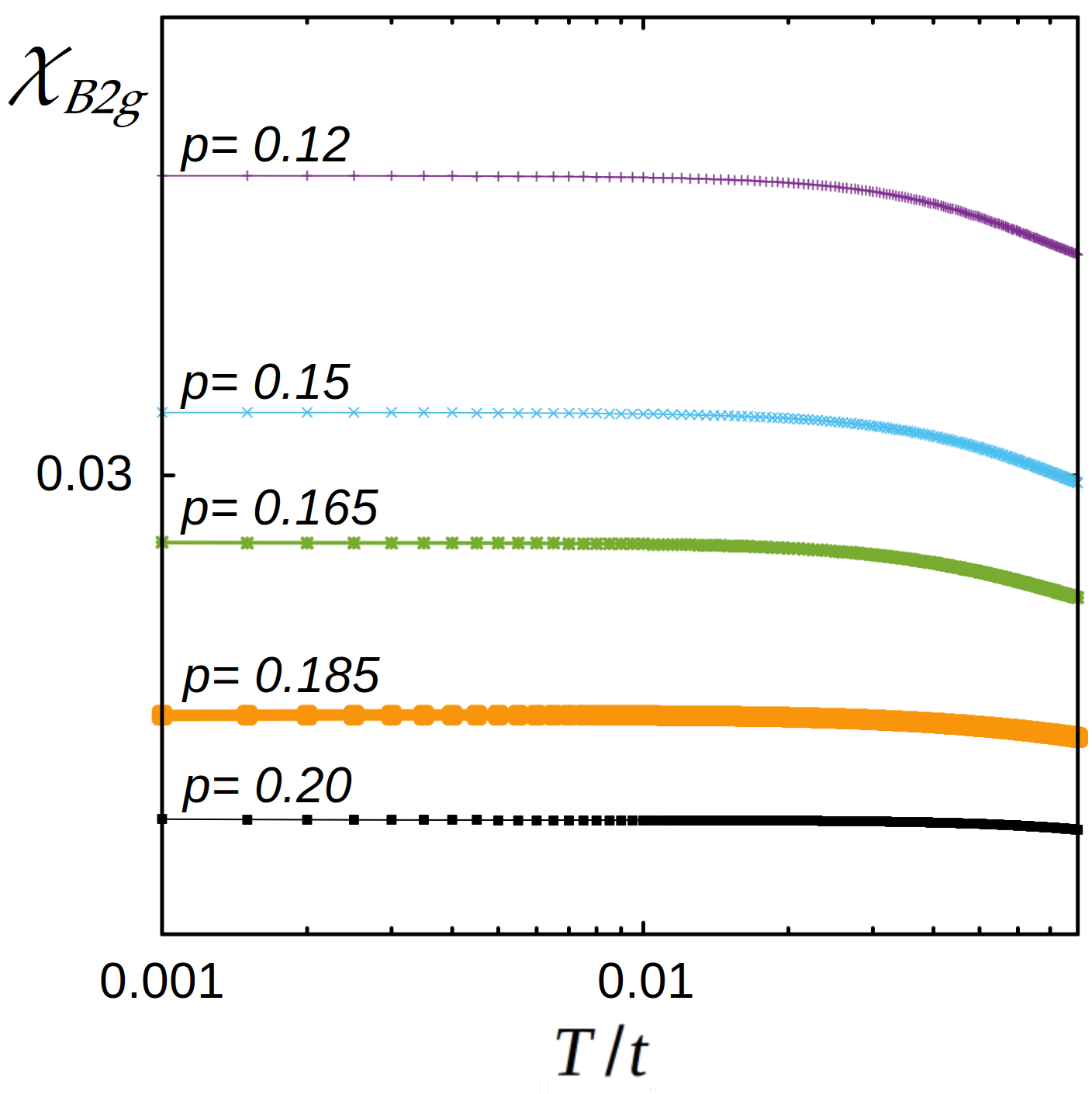}
\caption{(Color Online). Temperature dependence of the $\chi_{B_{2g}}(T)$ for various doping
$p$ across the exceptional van Hove singularity. }
\label{figS7}
\end{center}
\end{figure}

In summary, we show that ARPES spectra can detect the occurrence of an
exceptional van Hove Lifshitz transition at a doping $p = p_{ev}$ as a characteristic back-bending of
the quasiparticle dispersion. In the doping range $p^{*} > p > p_{ev}$ we expect to observe
a shoulder-like feature in the spectral function $A(\bk, \omega <0)$ close to $(\pi, 0)$,
which indicates the presence of an annular Fermi surface.
The persistence of the pseudogap in this doping range can be also detected if the unoccupied
states of the upper band $\omega_{1\mathbf{k}}$ are accessed. That would be a definite proof
that  $p_{ev} \neq p^{\ast}$, and that the Lifshitz transition and pseudogap end point are close but not coincident in the doping axis.

\subsection{Nematic susceptibilities}

In Fig.~\ref{figS6}(a) we show the temperature and doping dependence of the nematic susceptibility
$\chi_{B_{1g}}(T, p)$ on a logarithmic temperature scale.
For $p =p_{ev} \approx 0.185$, $\chi_{B_{1g}}(T)$
has considerable temperature dependence, while away from that doping the $T$-dependence is weak.
Likewise, Fig.~\ref{figS6}(b)
shows that the doping dependence of $\chi_{B_{1g}}(p)$ has a sharp peak around $p \sim p_{ev}$.
This is also the behavior seen in doped Bi$_2$Sr$_2$CaCu$_2$O$_{8+\delta}$. Moreover, above a
certain temperature cutoff, $\chi_{B_{1g}}(T)$ at the exceptional van Hove singularity has Curie-Weiss
$1/T$ scaling, as reported in the experiments, see Refs.~[5, 6] in the main text.

The above can be contrasted with that of $\chi_{B_{2g}}(T, p)$, the nematic susceptibility in the
$xy$ symmetry channel, for which the form factor is
$h_{\bk, B_{2g}} = \partial^2 \ep_{\bk}/\partial k_x \partial k_y$.
Since this symmetry channel probes the
nodal direction, it is mostly insensitive to the exceptional van Hove singularity, as shown
by the dashed blue line in Fig.~\ref{figS6}(b).
For the sake of completeness we show in Fig. \ref{figS7}
the temperature and doping dependence of $\chi_{B_{2g}}(T)$.
Unlike in the case of the $B_{1g}$ response,
the $B_{2g}$ response is mostly $T$-independent, even at $p_{ev}$. Furthermore,
since it is insensitive to the exceptional van Hove point $p_{ev}$, its
 behavior as a function of doping is monotonic.